# Photoacoustic-monitored laser treatment for tattoo removal: a feasibility study

YIYUN WANG[1,2,3], DAOHUAI JIANG[1,2,3], HENGRONG LAN[1,2,3], FENG GAO[1] AND FEI GAO[1,*]

[1]Hybrid Imaging System Laboratory, Shanghai Engineering Research Center of Intelligent Vision and Imaging, School of Information Science and Technology, ShanghaiTech University, Shanghai 201210, China
[2]Chinese Academy of Sciences, Shanghai Institute of Microsystem and Information Technology, Shanghai 200050, China
[3]University of Chinese Academy of Sciences, Beijing 100049, China

*gaofei@shanghaitech.edu.cn

**Abstract:** Skin blemishes and diseases have attracted increasing research interest in recent decades, due to their growing frequency of occurrence and the severity of related diseases. Various laser treatment approaches have been introduced for the alleviation and removal of skin pigmentation. The treatments' effects highly depend on the experience and prognosis of the relevant operators. But, the operation process lacks real-time feedback, which may directly reflect the extent of the treatment. In this letter, we report a photoacoustic-guided laser treatment method with a feasibility study, specifically for laser treatment targeting the tattoo's removal. The results well validated the feasibility of the proposed method through the experiments on phantoms and *ex vivo* pig skin samples.

## 1. Introduction

Skin pigmented lesions are more and more common nowadays. Such diseases may be caused by the innate factor, growing ultraviolet radiation due to the environmental and air pollution, excessive pressure and sleep insufficiency, and the natural ageing process that affects the skin with tissue degeneration [1]. Pigmented lesions, including solar lentigines, chloasma, melanocytic nevi, blue nevi and melasma [2], have particular aesthetic effects on people's daily life. Also, when the lesions develop into malignant skin diseases like malignant melanoma, such diseases' high malignancy and mortality draw people's increasing awareness for early intervention.

It is also noticed that the current laser treatments lack the overtreatment's monitoring or feedback system in the entire treating process. The therapeutic effect mostly depends on the therapists' experience, while the irreversible skin damage caused by overtreatment may be irrecoverable. The conventional assessment of the laser treatment is biopsy that is invasive and time-consuming. Photoacoustic detection and imaging, as a fast-developing and noninvasive sensing and imaging technology, have been studied on skin [3]–[8]. The resemblance between the laser treatment and the photoacoustic effect brings to our attention, which may be combined in an effective way.

This work shows a feasibility study of photoacoustic monitoring of the laser pigmentation removal. As a preliminary study, we focus on the epidermal layer of skin and select one kind of exogenous pigmentation, which is a tattoo, for convenience. At the aim of better monitoring the process, lower laser energy was used in the paper, comparing with the commonly selected energy of the commercial laser treatment devices. We first validate the treatment effect of the multi-pulse laser modality with lower energy through comparisons. Then, the feasibility of the photoacoustic monitoring of the laser therapy process is studied through phantom and *ex vivo* experiments.

## 2. Theory

The selective photo-thermolysis theory supports the laser therapies, describing the process of selective destruction through thermal necrosis and coagulation [1]. It states that the chromophore is selectively harmed if its heat-up time is shorter than its inherent thermal relaxation time, and the process does minimal harm to the surrounding tissue [10]. In this way, the pigment is broken into small pieces and vanishes through the macrophage activity and phagocytosis [11]. The process involves photo-disruption, which is also known as the optical breakdown. The breakdown occurs on the targeted soft tissue, so the blast wave generated in the treatment process cavitates. Thus, the cavitation mainly concerns mechanical effect, instead of thermal effect. We keep the laser's pulse repetition rate longer than the tissue's thermal relaxation time, which implies that the thermal effect is normally neglected in the process. The following reifies the process in the aspect of photoacoustic sensing.

The photoacoustic signal is excited when the pulsed laser is irradiated to the targeted tissue. According to the previous work, the initial pressure of generated PA wave could be described as [12]–[14]:

$$p_0 = \Gamma \eta_{th} \mu_a F_{pulse} \tag{1}$$

where $\Gamma$ denotes the Gruneisen parameter, $\eta_{th}$ is the conversion efficiency transforming optical absorption to pressure, $\mu_a$ describes the optical absorption coefficient, and $F_{pulse}$ denotes the light flux that is constant in this case. Since the nonlinear thermal effect is neglected, the Gruneisen parameter remains constant. During the process of tattoo removal, the pigment is scattered into dissociative segment pieces and fades in color (Fig. 1). The color variation results in decrease in the optical absorption coefficient, which directly affects the photoacoustic signal generation. Thus, such color variation could be monitored by photoacoustic sensing in real time.

When the overtreatment occurs, the skin is scalded and even burnt off. It leads to changes in the tissue's chemical compositions that results in certain variations of the inherent attributes. The detected signals' peak may thus descend in another form (e.g. increase a bit, then continue to decrease). At the aim of observing the whole process, the laser irradiation period is designed to be long enough, covering both the treatment and overtreatment stages.

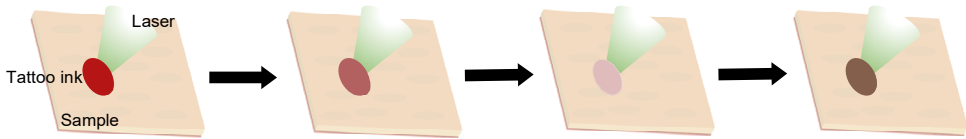

Fig. 1. Four stages of tattoo's color variation in the tattoo removal process. The color fades until the pigment is completely dispersed. When the laser continues irradiating the subject, the irradiated skin is scalded, which corresponds to the overtreatment.

## 3. Experiment

### 3.1 Sample Preparation

The experiment was designed to be conducted on a piece of tattooed pig skin since the pig skin can mimic human skin very well, compared with other animals' skins. The pig skin was tattooed with red lines by the professional. The conventional way for the professional to evaluate the treatment effect is the observation of the laser therapy based on personal experience. Firstly, we need to demonstrate the treatment consistency between a commercial laser device and our laser used in the experiment. We obtained the results through the commercial device available in clinics (PicoSure, CynoSure Inc.) as reference. As shown in Fig. 2, black box 1 circles the original tattooed area without treatment, and black box 2, 3 mark the regions of tattooed pigskins after the laser therapy.

The laser system used in our experiment is similar with the commercial laser treatment equipment, which is widely used in hospitals. To observe the complete tattoo-removing process, a relatively small laser output pulse energy was selected, which was 67 $mJ$. The other settings of our laser equipment were as equivalent as possible to the commercial ones. A 532 $nm$ wavelength pulsed laser (LPS-532-L, CNI Laser) with 5 $Hz$ repetition rate was chosen to conduct experiments. The tattoos in the black box 4 and 5 of Fig. 1 were irradiated by our laser system for about one minute. From observation in Fig. 1, the color variations in black box 4 and 5 are very similar to that in black box 2 and 3, which indicates that our laser system can achieve similar laser treatment effect as the conventional laser treatment equipment.

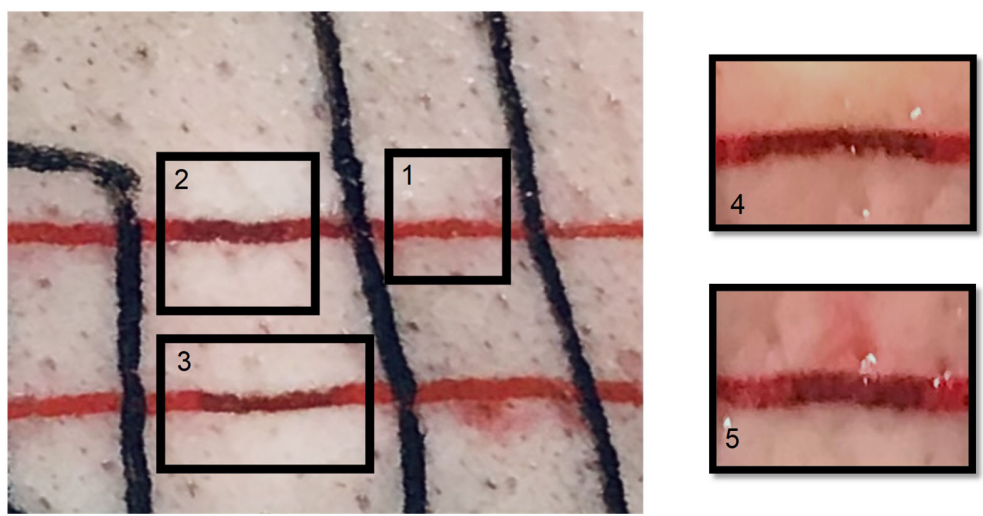

Fig. 2. A piece of pig skin with tattoos. The tattoo in black box 1 remained untreated. The tattoos in black box 2, 3 were undergone the first stage of the laser treatment using commercial laser treatment equipment. The tattoos in black box 4, 5 experienced the similar laser therapy using our laser system in the laboratory.

### 3.2 Experimental Setup

The experiment was conducted, using both phantom and pig skin samples. The first preliminary experiment was conducted on a white agar phantom that was tattooed with black ink (TI4002-5, Solong Tattoo). The pseudo tattoo on the phantom was mimicked through injecting a small amount of the ink from the side at a depth of about 2.4 mm.

The second and third *ex vivo* experimental sample was a piece of pig skin introduced above. Water acted as the ultrasonic transmission medium between the sample and the transducer for the first two experiments. For the third experiment, the transmission medium was changed into the ultrasonic coupling gel for better fitting the practical application scenarios.

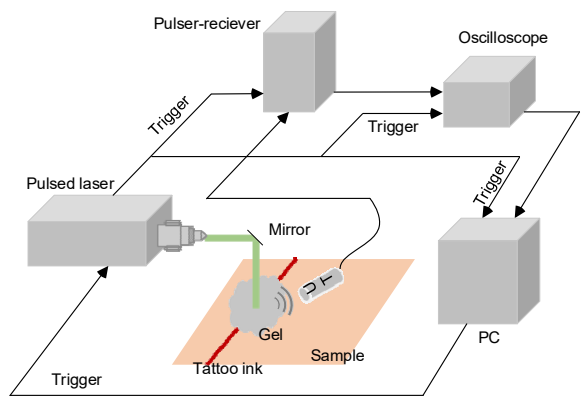

Fig. 3. Experimental setup of the photoacoustic monitoring tattoo-cleaning process system. UT, ultrasound transducer; PC, personal computer.

The experiment is designed to perform laser treatment for tattoo removal, collecting the photoacoustic signals excited from the targeted tissue simultaneously as feedback. As shown in Fig. 3, the excited photoacoustic signals were received by the ultrasound transducer (I5P6NF-H, Doppler Inc.) with 5 $MHz$ central frequency, and amplified with 46 $dB$ by the pulser-receiver (DPR300, JSR Ultrasonics Inc.). The pre-processed signals were then collected and averaged by the oscilloscope (DPO 5204B, Tektronix Inc.). The oscilloscope averaged the signals for 60 times to improve the signal-to-noise ratio.

*3.3 Results and Analysis*

The feasibility of the proposed method is firstly verified through a phantom experiment shown in Fig. 4(a)-(b). One typical time-domain waveform of the excited PA signals is shown in Fig. 4(c). The maximum peak that represents the signals' variations is chosen to plot along with the mimicking tattoo removal time, which is shown in Fig. 4(d). With the exponential function fitting, the fitting parameter $R^2$ achieves 95.19%. Fig. 4(d) clearly illustrates that the PA signal's peak amplitude descends as the increase of the irradiation time, which fits the proposed assumption very well. After about 70 seconds' irradiation, a hole occurred at the place of the irradiated spot in the middle of the circled tattooed phantom area in Fig. 4(b). This phenomenon represents over-irradiation. If the sample is a piece of human skin, the area would be burned away, which was assessed by the professional as over-treatment. The phantom experiment verified the feasibility of photoacoustic monitoring of the laser irradiation process.

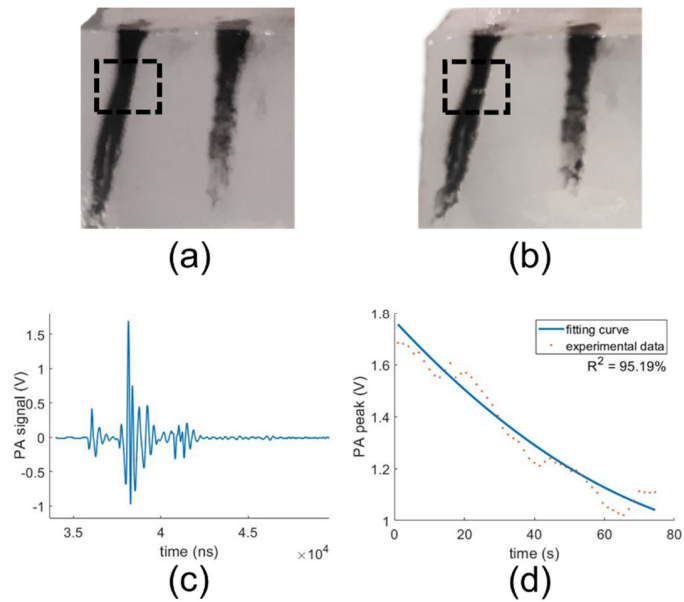

Fig. 4. (a) The phantom with mimicking tattoo. The dashed box marks the irradiated area. (b) The phantom after laser treatment. A hole appears in the middle of the dashed box. (c) A time-domain waveform of PA signal. (d) The chosen PA peak's amplitude variation during the tattoo removal process.

Aiming at reducing the possibility of overtreatment and optimizing the whole treatment process, we conducted the *ex vivo* experiment in two stages: the red-ink tattoo fading and the skin being scorched. The treatment performance in Fig. 5(a) and the reference performance in Fig. 1 have undistinguishable difference. We extracted the certain frequency components of the PA signals that mainly represent the second peaks of the original signals through wavelet decomposition. It is illustrated in Fig. 5(b) that the second peak may represent the signals' variation to the maximum extent due to the laser treatment. Of course, the six PA signals are chosen with an equal time interval. Through extracting the corresponding peaks of the original PA signals, the relationship between their amplitudes and the laser irradiation time is presented in Fig. 5(c). The general decreasing trend is consistent with the result in the former phantom experiment though the curve rises a bit, then turns flat in the middle of the treatment. The turning point at about 35 seconds of irradiation duration may indicate the complete pigment scattering. The distinctive amplitude variation from 35 to 50 seconds (duration B) may represent the unstable period between the pigment scattering stage and the skin-scorched stage. It is discovered that when the laser was irradiated to a piece of pure pigskin without any tattoo, the selected amplitude of PA signals presents descending tendency between 1.5~2V as shown in Fig. 5(d), which is quite similar with the curve between 50 to 70 seconds (duration C) in Fig. 5(c), indicating that it is the skin-scorched stage.

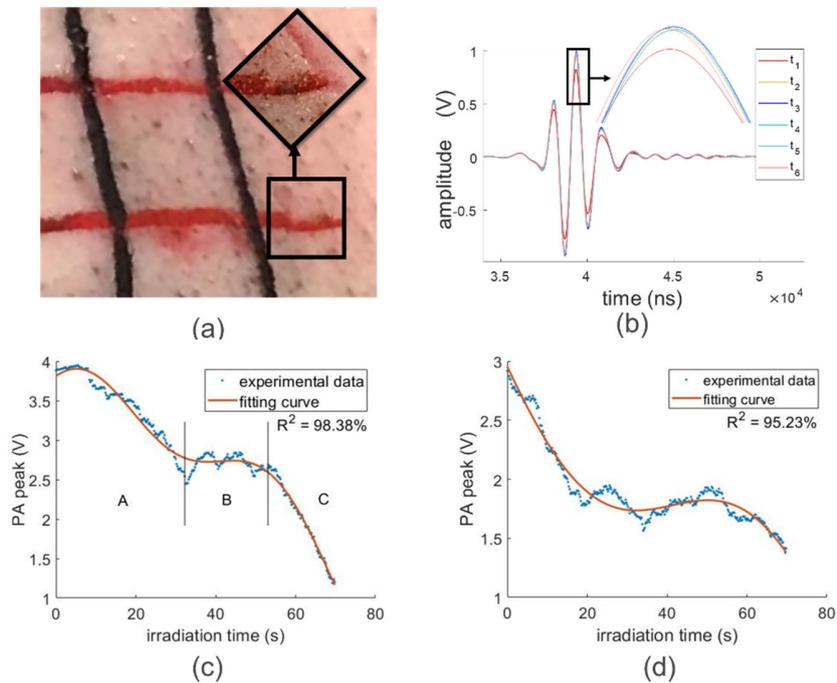

Fig. 5. (a) The marked area's change in colour on a piece of red-tattooed pigskin before and after a period of irradiation. (b) The selected amplitudes of six time-domain PA signals are chosen at the equal time interval, and the peaks are magnified. The waveforms are processed through wavelet decomposition. (c) The amplitude variation of the selected PA peaks during the process that the laser irradiates the tattoo area. Duration A: pigment scattering, duration B: PA peak signal's oscillation, duration C: skin being scorched. (d) The amplitude variation of the selected PA peaks with laser irradiating a piece of untattooed pigskin.

To better mimic the practical scenario, the third experiment was conducted with the ultrasonic coupling gel as the transmission medium. Using the same peak selection method for the PA signals shown in Fig. 6(a), we located the selected peak and derived the correlation. Unlike above experiment, the ink colour variation of the tattooed pigskin is hard to be identified with the naked eyes after 70 seconds' irradiation. However, when the cotton swab was used to dab the surface of the irradiated area, some of the red ink remained on the swab. This indicates that only the pigments' scattering exists during the process. From Fig. 6(b), it is suspected that the pigment was already scattered after about 30 seconds of irradiation. Due to the *ex vivo* feature of this experiment, metabolism no longer exists in the excised pigskin, which may result in the incapability of diffusion of the scattered pigments that presents as the curve turning flat. Another probable factor is the uneven surface of the coupling gel. The unevenness of the surface may vary the laser's refractive index that causes tiny variances in location and size of the light spot. It might reduce the laser's energy density, resulting in the indistinguishable colour change of the tattoo.

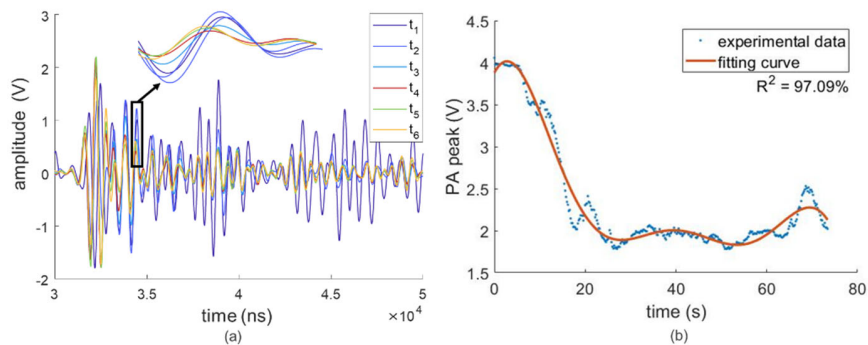

Fig. 6. (a) The chosen amplitudes of PA signals selected at an equal time interval, which are the selected waveforms after wavelet decomposition of the original PA signals. (b) The amplitude variation of the selected PA peaks during the laser treatment process.

## 4. Conclusion

To sum up, a proposed method of photoacoustic-monitored laser treatment for the tattoo removal process was demonstrated in this paper. Three experiments illustrate the feasibility of the proposed PA-monitored laser treatment method step by step, and reveal the relationship between the amplitude of the selected PA signals' peak and the laser treatment period. With laser irradiation targeting the pigskin sample, the PA amplitude first descends with the ascending laser irradiation duration in the process of pigment scattering. Then, it remains flat with oscillation. If the overtreatment occurs, the amplitude may trend downward again. The photoacoustic detection method shows its unique ability to monitor the entire laser treatment process in different stages. In the future work, we will further optimize the system with more precise laser treatment monitoring and more integrated system development, and will explore *in vivo* applications.